\documentclass[sigconf]{aamas}  

\usepackage{booktabs}

\setcopyright{ifaamas}  
\acmDOI{doi}  
\acmISBN{}  
\acmConference[AAMAS'18]{Proc.\@ of the 17th International Conference on Autonomous Agents and Multiagent Systems (AAMAS 2018), M.~Dastani, G.~Sukthankar, E.~Andre, S.~Koenig (eds.)}{July 2018}{Stockholm, Sweden}  
\acmYear{2018}  
\copyrightyear{2018}  
\acmPrice{}  

\usepackage[most]{tcolorbox}
\usepackage{ragged2e}
\usepackage{tikz}
\usepackage{graphicx}
\usepackage{pgfplots}
\pgfplotsset{compat=1.14}
\usepackage{soul}
\usepackage{longtable}
\usepackage{color, colortbl}

\usepackage{supertabular,booktabs}
\usepackage{array,tabularx}
\usepackage{flushend}
\newcolumntype{C}[1]{>{\centering\arraybackslash}m{#1}}
\newcolumntype{Y}{>{\centering\arraybackslash}X}

\begin{document}

\title{Learning Optimal Redistribution Mechanisms Through Neural Networks}  



\author{P Manisha}
\affiliation{%
  \institution{International Institute of Information Technology, Hyderabad, India.}
}
\email{manisha.padala@research.iiit.ac.in}

\author{C V Jawahar}
\affiliation{%
  \institution{International Institute of Information Technology, Hyderabad, India.}
}
\email{jawahar@iiit.ac.in}

\author{Sujit Gujar}
\affiliation{%
  \institution{International Institute of Information Technology, Hyderabad, India.}
}
\email{sujit.gujar@iiit.ac.in}

\begin{abstract}  
We consider a social setting where $p$ public resources/objects are to be allocated among $n$ competing and strategic agents so as to maximize social welfare (the objects should be allocated to those who value them the most). This is called allocative efficiency (AE). We need the agents to report their valuations for obtaining these resources, truthfully referred to as dominant strategy incentive compatibility (DSIC). Typically, we use auction-based mechanisms to achieve AE and DSIC. However, due to Green-Laffont Impossibility Theorem, we cannot ensure budget balance in the system while ensuring AE and DSIC. That is, the net transfer of money cannot be zero. This problem has been addressed by designing a redistribution mechanism so as to ensure minimum surplus of money as well as AE and DSIC. The objective could be to minimize surplus in expectation or in the worst case and these $p$ objects could be homogeneous or heterogeneous. Designing of such mechanisms is non-trivial. Especially, designing redistribution mechanisms which perform well in expectation becomes analytically challenging for heterogeneous settings. 

In this paper, we take a completely different, data-driven approach. We train a neural network to determine an optimal redistribution mechanism based on given settings with both the objectives, optimal in expectation and optimal in the worst case. We also propose a loss function to train a neural network to optimize worst case. We design neural networks with the underlying rebate functions being linear as well as nonlinear in terms of bids of the agents. Our networks' performances are same as the theoretical guarantees for the cases where it has been solved. We observe that a neural network based redistribution mechanism for homogeneous settings which uses nonlinear rebate functions outperforms linear rebate functions when the objective is optimal in expectation. Our approach also yields an optimal in expectation redistribution mechanism for heterogeneous settings.

\end{abstract}

%

\keywords{VCG Mechanisms; Dominant Strategy Incentive Compatibility (DSIC); Redistribution Mechanism; Neural Networks; Deep Learning}  

\maketitle


\section{Introduction}
\label{sec:intro}In this paper, we address the problem of allocating public resources/objects among multiple agents who desire them. These strategic agents have their private values for obtaining the resources. The allocation of the objects should be such that the society, as a whole, gets the maximum benefit. That is, the agents who value these resources the most should get them. This condition is referred to as \emph{Allocative Efficient (AE)}. To achieve this, we need the true valuations of the agents, which the strategic agents may misreport for personal benefit. Thus, there is a need for an auction-based mechanism.  A mechanism ensuring truthful reporting, is called \emph{Dominant Strategy Incentive Compatible} (DSIC).  The classical Groves mechanisms \cite{Groves73} satisfy both of these properties. 

Groves mechanisms achieve DSIC by charging each agent an appropriate amount of money known as Groves' payment rule. The most popular among Groves mechanism is VCG mechanism \citep{Vickrey61,Clarke71,Groves73}.  Use of VCG mechanism results in collection of money from the agents. It should be noted that our primary motive in charging the agents is to elicit their true valuations and not to make money from them as the objects are public. Hence, we need to look for the other Groves mechanisms. Moreover, the mechanism cannot fund the agents. Thus, we desire a mechanism that incurs neither deficit nor surplus of funds; it must be \emph{Strictly Budget Balanced} (SBB).  However, due to Green-Laffont Impossibility Theorem \cite{Green81}, no mechanism can satisfy AE, DSIC, and SBB simultaneously.  Thus, any Groves payment rule always results in either surplus or deficit of funds. 

To deal with such a situation, Maskin and Laffont \cite{LaffontM79} suggested that we first execute VCG mechanism and then redistribute the surplus among the agents in a manner that does not violate DSIC. This mechanism is referred to as a \emph{Groves' redistribution mechanism} or simply \emph{redistribution mechanism} (RM) \cite{Gujar11,Guo12} and the money returned to an agent is called it's \emph{rebate}. The rebates are determined through \emph{rebate functions}. Thus, designing a RM is same as designing an underlying rebate function. 

In the last decade, a lot of research focused on dealing with the Green-Laffont Impossibility theorem and on designing an optimal redistribution mechanism (RM) that ensures maximum possible total rebate \cite{Faltings05,cavalloRedis06,Guo07,Guo08,Gujar11,Clippel14,Guo12,Guo11, Gul99}.  An optimal RM could be optimal in expectation or optimal in the worst case. The authors of \cite{Guo07,Guo09,Moulin09} address the problem of finding the optimal RM when all the objects are identical (homogeneous). Guo and Conitzer \cite{Guo09,GuoOEL08} model an optimization problem and solve for an optimal linear rebate function which guarantees maximum rebate in the worst-case (WCO) and optimal in expectation for a homogeneous setting. The authors of \cite{Gujar11,Guo12} extend WCO to a heterogeneous setting where the objects are different and propose a nonlinear rebate function called as \emph{HETERO}. Despite HETERO being proved to be optimal for unit demand in heterogeneous settings, in general, it is challenging to come up with a nonlinear RM analytically. Analytical solutions for an optimal in expectation RMs in heterogeneous settings is elusive. Moreover, the possibility of a nonlinear rebate function which is optimal in expectation for homogeneous setting has not been explored yet. Thus, there is a need for a new approach towards designing RMs. In this paper, we propose to use neural networks and validate its usefulness.

Neural networks have been successful in learning complex, nonlinear functions accurately, given adequate data \citep{hornik91}. There is a theoretical result which states a neural network can approximate any continuous function on compact subspace of $\mathbb{R}^n$ \cite{hornik89}. However, designing such a network has been elusive until recent times. The latest theoretical developments ensure that stochastic gradient descent (SGD) converges to globally optimal solutions \cite{Kawaguchi16,daniel16}. In recent times, with advent in computing technology, neural networks have become one of the most widely used learning models. They have outperformed many of the traditional models in the tasks of classification and generation etc. \cite{kaiming16, Graves13}. In the context of game theoretic mechanism design, D{\"{u}}tting \emph{et.al.} \cite{optAuctions} have proposed neural network architectures for designing optimal auctions. Converse to ours', their goal is to design the network for maximizing the expected revenue. Tang \cite{tang17} uses deep reinforcement learning for optimizing a mechanism in dynamic environments. Their goal, unlike ours is not to design a mechanism, but to find the optimal parameters for the existing mechanism. Our goal in this paper is to study: Can we train neural networks to learn optimal RMs with the help of randomly generated valuation profiles, rather than analytically designing them? 
The following are our contributions.

\paragraph{Contributions}
To the best of our knowledge, this paper is a first attempt towards learning optimal redistribution mechanisms (RM) using neural networks. 
\begin{itemize}
\item To begin with, we train neural networks for the settings where researchers have designed RMs analytically. In particular, we train networks, OE-HO-L and OW-HO-L for optimal in expectation for homogeneous settings with linear rebate functions and optimal in the worst case for homogeneous settings with linear rebate functions respectively. Both the neural networks match the performance of theoretically optimal RMs for their respective settings. 
\item Next, we train a network, OW-HE-NL to model nonlinear rebate function for optimal in worst case RM in heterogeneous settings, discarding the need to solve it analytically.  Note that, traditionally, neural networks have been mostly used for stochastic approximation of an expectation, but our model is also able to learn a worst-case optimal RM as well.
\item Motivated by the network performance in above, we train OE-HO-NL, an optimal in expectation RM with nonlinear rebate function for the homogeneous setting. We find that this model ensures greater expected rebate than the optimal in expectation RM with linear functions, proposed by Guo and Conitzer \cite{GuoOEL08}.
\item  We also train OE-HE-NL, an optimal in expectation for heterogeneous setting with nonlinear functions and we experimentally observe that it's performance is reasonable.
\end{itemize}

\paragraph{Organization}
In Section \ref{sec:rel_work}, we describe the related work. In Section \ref{sec:prelims}, we explain the notation used in the paper and preliminaries of
Groves redistribution mechanism. We develop our neural network approach in Section
\ref{sec:nn}. We discuss the training of neural networks and the experimental 
analysis in Section \ref{sec:expr}. We conclude the paper with the possible future directions in Section \ref{sec:con}.

\section{Related Work}
\label{sec:rel_work}
In this section, we discuss the research related to our work. 
\paragraph{Auction Based Mechanism}
To obtain the required private information from strategic agents truthfully, \emph{mechanism design theory} is developed \cite{nisan07,Garg08a}. The key idea is to charge the agent appropriately to make mechanisms truthful or DSIC. The most popular auction-based mechanisms are VCG and Groves mechanisms \cite{Vickrey61, Clarke71, Groves73} which satisfy the desirable properties, namely, allocative efficiency (AE) and dominant strategy incentive compatibility (DSIC).  Another desirable property is the net transfer of the money in the system should be zero, i.e., it should be \emph{strictly budget balanced} (SBB). Green and Laffont \cite{ Green81} showed no mechanism can satisfy AE, DSIC, and SBB simultaneously.  As we cannot compromise on DSIC, we must compromise on one of AE or SBB. 

Faltings \cite{Faltings05} and Guo and Conitzer \cite{Guo08} achieved budget balance by compromising on AE. Hartline and Roughgarden \cite{Hartline08} proposed a mechanism that maximizes the sum of the agents' utilities in-expectation. Clippel \emph{et. al.} \cite{Clippel14} used the idea of destroying some of the items to maximize the agents' utilities leading to approximately AE and approximately SBB. A completely orthogonal approach was proposed by Parkes \emph{et. al.} \cite{parkes01}, where the authors propose an optimization problem which is approximately AE, SBB and though not DSIC, it is not easy to manipulate the mechanism.   However, an aggressively researched approach is to retain AE and DSIC and design mechanism that is as close to SBB as possible. These are called \emph{redistribution mechanisms} (RM). 

\paragraph{Redistribution Mechanisms}  Maskin and Laffont \cite{LaffontM79} first proposed the idea of redistribution of the surplus as far as possible after preserving DSIC and AE.  Bailey \cite{Bailey97}, Cavallo \cite{cavalloRedis06}, Moulin \cite{Moulin09}, and Guo and Conitzer \cite{Guo07} considered a setting of allocating $p$ homogeneous objects among $n$ competing agents with unit demand. and Guo and Conitzer \cite{Guo09} generalized their work in \cite{Guo07} to multi-unit demand to obtain worst-case optimal (WCO) RM. In \cite{GuoOEL08}, Guo and Conitzer designed RM that is optimal in expectation for homogeneous settings. 

Gujar and Narahari \cite{Gujar11} proved that no linear RM can assure non-zero rebate in worst case and then generalized  WCO mechanism mentioned above  to heterogeneous items, namely HETERO. Their conjecture that HETERO was feasible and worst-case optimal was proved by Guo \cite{Guo12} for heterogeneous setting with unit-demand. 

\section{Preliminaries}
\label{sec:prelims} 
Let us consider a setting comprising $p$ public resources/objects and $n$ competing agents who assign a certain valuation to these objects. Each agent desires at most one  out of these $p$ objects. These objects could be homogeneous, in which case, agent $i$ has valuation $v_i=\theta_i$ for obtaining any of these $p$ resources. It could also be the case that the objects are distinct or heterogeneous and each agent derives different valuation for obtaining different objects ($v_i= ({\theta_i}_1,{\theta_i}_2,\ldots,{\theta_i}_p)$). These objects are to be assigned to those who value it the most, that is, it should be allocatively efficient (AE). The true values $\mathbf{v}=(v_1,v_2\ldots,v_n)$ that the agents have for the objects are based on their private information $\theta=(\theta_1,\theta_2,\ldots,\theta_n)=(\theta_i,\theta_{-i})$ and the strategic agents may report them as $(\theta_1',\ldots,\theta_n')$. In the absence of appropriate payments, the agents may boast their valuations. 
Hence, we charge agent $i$ a payment $m_i(\theta')$ based on the reported valuations. 

We need to design a mechanism $\mathcal{M}=(\mathcal{A},\mathcal{P})$, an allocation rule $\mathcal{A}$ and a payment rule $\mathcal{P}$. $\mathcal{A}$ selects an allocation $k(\theta')\in K$ where $K$ is the set of all feasible allocation and $\mathcal{P}$ determines the payments.  With this notation, we now explain the desirable properties of a mechanism.
\subsection{Desirable Properties}
\label{subsec:prop}
One of our primary goals is to ensure allocative efficiency. 
\begin{definition}[Allocative efficiency (AE)]- A mechanism $\mathcal{M}$ is \emph{allocatively efficient} (AE) if it choses in every given type profile, an allocation of objects among the agents such that sum of the valuations of the allocated agents is maximized. That is, for each $\theta \in \Theta$,
$$k^*(\theta) \mbox{ } \in \mbox{ } \underset{k\in K}{\operatorname{argmax}} \displaystyle \sum_{i=1}^nv_{i}(k, \theta_{i}).$$
\end{definition}

The most desirable property is that the agents should report their valuations truthfully to the mechanism. Formally, it is called \emph{dominant strategy incentive compatibility} (DSIC).
\begin{definition}[Dominant Strategy Incentive Compatibility (DSIC):] We say a mechanism $\mathcal{M}$ to be \emph{dominant strategy incentive compatible} (DSIC), if it is a best response for each agent to report their type truthfully, irrespective of the types reported by the other agents. That is, 
$$v_{i}(k(\theta_{i}, \theta_{-i}),\theta_{i}) - m_i(\theta_{i}, \theta_{-i}) \geq v_{i}(k(\theta_{i}', \theta_{-i}),\theta_{i}) - m_i(\theta_{i}', \theta_{-i})$$
$$\forall \theta_{i}' \in \Theta_{i}, \forall \theta_{i}\in\Theta_{i}, \forall \theta_{-i} \in \Theta_{-i}, \forall i \in N. $$
\end{definition}

Given an AE allocation rule, Groves proposed a class of mechanisms known as \emph{Groves' mechanisms} that ensure DSIC. Groves' payment rule is $m_i(\theta) = - \displaystyle \sum_{j \neq i } v_{j}(k_{*}(\theta),\theta_{j}) + h_i(\theta_{-i})$, where $h_i$ is an arbitrary function of reported valuations of the agents other than $i$. 
Clarke's payment rule is a special case of Groves' payment rule where $h_i(\theta_{-i})= \displaystyle \sum_{j \neq i } v_{j}(k_{-i}^*(\theta_{-i}),\theta_{j}) $ where $k_{-i}^*$ is an AE allocation when the agent $i$ is not part of the system. Thus, Clarke's payment rule is give by \Cref{eq:v_pay}
\begin{equation}
\label{eq:v_pay}
t_{i}(\theta) = \displaystyle \sum_{j \neq i } v_{j}(k_{-i}^*(\theta_{-i}),\theta_{j})  - \displaystyle \sum_{j \neq i } v_{j}(k^{*}(\theta),\theta_{j}) \mbox{ } \forall i = 1,\ldots,n 
\end{equation}
And, $m_{i} = t_{i}$. This payment scheme is referred to as \emph{VCG payment}. The total payment by all the agents is, $t(\theta) = \sum_{i \in N}t_{i}(\theta)$

Another property we desire is budget balance condition. 
\begin{definition}[Budget Balance condition or Strictly Budget balance (SBB)] We say that a mechanism  $\mathcal{M}$ is \emph{strictly budget balanced} (SBB) if for each $\theta \in \Theta, m_{1}(), m_{2}(),\ldots,m_{n}()$ satisfy the condition, $ \sum_{i\in N} m_{i}(\theta) = 0$
\end{definition} 
It is weakly budget balanced if $ \sum_{i\in N} m_{i}(\theta) \geq 0.$
\justify

One can implement AE allocation rule and charge the agents VCG payments. In the case of auctions settings, the seller collects the payments. In our setting, the goal is not to make money as these objects are public resources. However, in general, due to Green-Laffont Impossibility Theorem \cite{Green81}, no AE and DSIC mechanism can be strictly budget balanced. That is, the total transfer of money in the system may not be zero. So, the system will be either left with a surplus or incur deficit. Using Clarke's mechanism, we can ensure under fairly weak conditions, that there is no deficit of money (that is, the mechanism is weakly budget balanced) \cite{Clarke71}. The idea proposed by Maskin and Laffont \cite{LaffontM79} is to design a payment rule as to first collect VCG payments and then redistribute this surplus (rebate) among the agents while ensuring DSIC. This leads to \emph{Groves' Redistribution Mechanism}, in which the rebate is given by a rebate function $r()$.

\subsection{Groves' Redistribution Mechanism} Since SBB cannot coexist with DSIC and AE, we would like to redistribute the surplus to the participants as much as possible, preserving DSIC and AE. Such a mechanism is referred to as \textit{Groves redistribution mechanism} or simply \textit{redistribution mechanism}. Designing a redistribution mechanism involves designing an appropriate rebate function. We desire a rebate function which ensures maximum rebate (which is equivalent to minimum budget imbalance). In addition to DSIC, we want the redistribution mechanism to have following properties:  
\begin{enumerate}
	\item Feasibility (F): the total payment to the agents should be less than or equal to the total received
		payment. 
	\item Individual Rationality (IR): each
		agent's utility by participating in the mechanism
		should be non-negative.
	\item Anonymity: rebate function is same for all the agents, $r_{i}() = r_{j}() = r()$.  This may still result is different redistribution payments as the input to the function may be very different. 
\end{enumerate}
While designing redistribution mechanism for either homogeneous or heterogeneous objects,  we may have linear or nonlinear rebate function of the following form,

\begin{theorem} \cite{Gujar11} \label{thm:order} In the Groves redistribution mechanism, any deterministic, anonymous rebate function $f$ is DSIC iff, 
$$r_{i} = f(v_{1},v_{2},\ldots,v_{i-1},v_{i+1},\ldots,v_{n})	 \mbox{ } \forall i \in N$$
where, $v_{1} \geq v_{2} \geq \ldots \geq v_{n}.$
\end{theorem}
 
\begin{definition}\label{def:lr}[Linear Rebate Function] The rebates to an agent follow a \emph{linear rebate} function if the rebate is a linear combination of bid vectors of all the remaining agents. Thus, $r_{i}(\theta, i) = c_0 + c_1v_{-i,1} +\ldots+ c_{n-1}v_{-i,n-1}.$ \end{definition}

There may exist a family of redistribution mechanisms which satisfy the above constraints, but the aim is to identify the one mechanism that redistributes the greatest fraction of the total VCG payment. To measure the performance of redistribution mechanism \cite{Guo09}, defines redistribution index, 
\begin{definition} \label{def:RI}[Redistribution Index] The \emph{redistribution index} of a redistribution mechanism is defined to be the worst case fraction of VCG surplus that gets redistributed among the agents. That is, $$e^{ow} = \displaystyle inf_{\theta:t(\theta)\neq 0} \frac{\sum r_{i}(\theta_{-i})}{t(\theta)} $$
\end{definition}
With the notation defined above and Green-Laffont Impossibility theorem in backdrop, we now explain existing redistribution mechanisms in the following subsection.

\subsection{Optimal Redistribution Mechanisms}
\label{subsec:opt}
It may happen that, one mechanism might redistribute higher rebate at $\theta_{1}$ and another mechanism at $\theta_{2}$. Hence, we use the two kinds of evaluation metrics defined to select a mechanism. One metric compares the rebate functions based on maximum expected total redistribution, to find the mechanism which is optimal in expectation. The other metric finds the optimal in worst-case redistribution mechanism, based on the lowest redistribution index it guarantees
\subsubsection{Optimal in Expectation} (OE): If the prior distributions over agents' valuations are available we can compare the mechanisms based on total expected redistribution. In \cite{GuoOEL08} the authors derive the mechanism and prove its optimality for homogeneous setting with linear rebate function. We define a redistribution index for OE setting as follows: $$e^{oe} = \frac{\mathbb{E}{\sum_{i}r_{i}(\theta_{-i})}}{\mathbb{E}{\sum_{\theta}t(\theta)}}$$ Maximizing $e^{oe}$ is equivalent to maximizing the expected total rebate. The authors formulated the problem as given in \Cref{tab:opt}.

\begin{table}
\caption{Optimization problem formulation}
\label{tab:opt}
\begin{tabular}{|l|l|l|}
	\hline
	 	& OE & OW \\
	        \hline
	        Variables : & $c_0$, $c_1$,....,$c_{n-1}$ & $e^{ow}, c_0$, $c_1$,....,$c_{n-1}$ \\
      
	        Maximize : & $\mathbb{E}{\sum_{i=1}^n r_{i}}$ & $e^{ow}$ \\
                  \hline
	        Feasibility : & $\sum_{i=1}^n r_{i}$ <= $t$  & $\sum_{i=1}^n r_{i} \leq t$ \\
                  \hline
	        Other constraints & & \\
            For worst-case :  &  & $\sum_{i=1}^n r_{i} \geq e^{ow}t$ \\
	        IR :&  & $r_{i} \geq 0$   \\
	        \hline
	\end{tabular}
\end{table}

The OE objective for heterogeneous objects as well as with nonlinear rebate function has not been addressed yet.
\subsubsection{Optimal in Worst-case} (OW) \label{subsec:wco} : The redistribution mechanism is better if it ensures higher rebate to the agents on average. In the absence of distributional information, we would evaluate a mechanism by considering the worst redistribution index that it guarantees. In \cite{Guo09} the authors gave the following model and analytically solved it for homogeneous setting with linear rebate functions. They also claim the worst-case optimal mechanism is optimal among all redistribution mechanisms that are deterministic, anonymous and satisfy DSIC, AE and F. The optimization problem is formulated as given in \Cref{tab:opt}.

For heterogeneous setting, \cite{Gujar11} defines a nonlinear redistribution mechanism which is called HETERO and \cite{Guo12} proves the optimality of HETERO. There is no optimal mechanism with linear rebate function for heterogeneous setting as established by the following theorem, 
\begin{theorem}\cite{Gujar11}
\label{thm:imp}
If a redistribution mechanism is feasible and individually rational, then there cannot exist a linear rebate function which is simultaneously DSIC, deterministic, anonymous and has non-zero redistribution index.  
\end{theorem} 
Equipped with the knowledge of the existing approaches and above theorem, we describe our approach for designing the optimal rebate function using neural networks. 

\subsection{Our Approach}
As mentioned, for a homogeneous setting the linear rebate functions that are OE and OW can be analytically found by formulating redistribution mechanism as a linear program. However, for heterogeneous settings, linear redistribution mechanism need not be a good choice (\Cref{thm:imp}). Even though \cite{Guo12} has solved for redistribution mechanisms in heterogeneous by proving HETERO to be OW,  an OE mechanism for heterogeneous settings has not been formulated yet. Moreover, OW mechanism (HETERO) is not simple to describe. In addition, for homogeneous settings, it is not known whether nonlinear redistribution mechanisms can do better than linear for OE objective.

In this paper, we address these issues, using a novel data-driven approach to approximate a rebate function for a given setting, without analytically solving for it. That is, we generate a large number of bid profiles randomly and train a neural network to determine the rebates for the agents so as to achieve the given objective, either OE or OW. We consider both the rebate functions, linear and nonlinear for homogeneous objects as well as heterogeneous objects. The choice of neural networks is largely motivated by the universal approximation theorem, which states that a feed-forward network with single hidden layer containing a finite number of neurons can approximate continuous functions on compact subsets. 
\section{Design of Neural Networks}
\label{sec:nn}
Neural Networks are biologically inspired paradigms which learn optimal functions from data. The main components that customize such a network for a specific task are its architecture and the objective function which guides its training. To design a rebate function which is OE or OW, we define an appropriate neural network in \Cref{subsec:NN}. We begin by describing an artificial neuron which is the fundamental processing unit of a neural network. 
\subsection{Basic Structure} 
\label{subsec:bs}
An artificial neuron receives inputs $x_{0}\mbox{ }to\mbox{ }x_{m}$. If necessary, we apply nonlinear activation function $\phi$ to obtain the output ($y$) from the neuron. The activation function is for thresholding the output to introduce nonlinearity in the network. Thus the output of a neuron is, 
$$y = \phi (\sum_{i=0}^{m} w_{i}x_{i})$$ where $w_{i}$ is the weight for $i^{th}$ input.\\
In any general network, we connect the neurons together in a specific and useful manner. The neurons are grouped into primarily three layers, input layer, hidden layer, and output layer. The weights are randomly initialized before training. Given a set of training input and output pairs, the model compares it's own output with the desired output and tries to learn the optimal set of weights by back-propagating the error through the network. In our case, we do not have a desired output, but we have an objective function that is to be maximized. That is, we need to determine optimal weights such that the rebate function is OE or OW. In addition, our mechanism should be Feasible and Individual Rational. The total VCG payment by the agents is $t$, and the neural network parameters are $(w,b)$ then,  
\begin{itemize}
\item \label{cond:F} Feasibility : $g(w,b) : t - \displaystyle \sum_{i=1}^n r_{i}(w,b) \geq 0$ 
\item \label{cond:IR} Individual Rationality : $g'(w,b) : r_{i}(w,b) \geq 0, \mbox{ }\forall i \in N$
\end{itemize}
The above inequality constraints are added to the loss function during training. Having defined a general network and the constraints, we define the specific design of the network that we use. 
\subsection{Neural Network Architecture} 
\label{subsec:NN}
\subsubsection{Linear rebate function :}
To model the linear rebate function as given by \Cref{def:lr}, we use a network consisting of neurons with $n$ input and $n$ output nodes without any activation function. The input nodes represent agents' valuations and output nodes represent their rebate. As required by \Cref{thm:order}, a rebate function for an agent $i$ should depend only on valuations of the remaining agents. Hence, we connect the $i^{th}$ output node to all the input nodes except $i^{th}$ input node as shown in \Cref{fig:linear}. We used a total of $n-1$ weights and $1$ bias. Since the weights and the bias used is same for calculating the rebate of each agent (represented by each node in the output layer) we ensure that the $r()$ is anonymous. In addition to the weights ($w$) which model the $c_{1}\mbox{ }to\mbox{ }c_{n}$ in $r()$, there is a same bias added ($b$), which models the $c_0$, to each output.\\ 
\centerline{$r_i$ = $\displaystyle \sum_{j=1}^{n-1} v_{i}w_{j} + b $,    $\forall i = 1 \rightarrow n$}

\tikzset{%
   neuron missing/.style={
    draw=none, 
    scale=4,
    text height=0.333cm,
    execute at begin node=\color{black}$\vdots$
  },
}
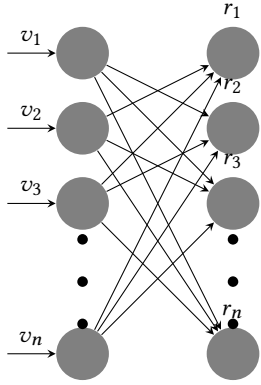
\begin{figure}
\centering
\begin{tikzpicture}[ >=stealth]

\foreach \m/\l [count=\y] in {1,2,3}
{
 \node [circle,fill=black!50,minimum size=0.7cm] (input-\m) at (0,2.5-\y) {};
}
\foreach \m/\l [count=\y] in {4}
{
 \node [circle,fill=black!50,minimum size=0.7cm ] (input-\m) at (0,-2.5) {};
}
 
 \node [neuron missing]  at (0,-1.5) {};
 
\foreach \m/\l [count=\y] in {1,2,3}
{
 \node [circle,fill=black!50,minimum size=0.7cm] (hidden-\m) at (2,2.5-\y) {};
}
\foreach \m/\l [count=\y] in {4}
{
 \node [circle,fill=black!50,minimum size=0.7cm ] (hidden-\m) at (2,-2.5) {};
}
 
 \node [neuron missing]  at (2,-1.5) {};

\foreach \l [count=\i] in {1,2,3,n}
  \draw [<-] (input-\i) -- ++(-1,0)
    node [above, midway] {$v_{\l}$};
    
\foreach \l [count=\i] in {1,2,3,n}
  \node [above] at (hidden-\i.north) {$r_{\l}$};

\foreach \i in {1}
  \foreach \j in {2,...,4}
    \draw [<-] (hidden-\i) -- (input-\j);

\foreach \i in {2}
  \foreach \j in {1,3,4}
    \draw [<-] (hidden-\i) -- (input-\j);

\foreach \i in {3}
  \foreach \j in {1,2,4}
    \draw [<-] (hidden-\i) -- (input-\j);

\foreach \i in {4}
  \foreach \j in {1,2,3}
    \draw [<-] (hidden-\i) -- (input-\j);

\end{tikzpicture}
\caption{\label{fig:linear} Network model}
\end{figure}

\subsubsection{Nonlinear rebate function :} 
The network consists of neurons with $n$ input and $n$ output nodes and one hidden layer. The input nodes represent agent valuations and output nodes represent the rebate. The neurons in the input and hidden layer use ReLU activation which returns $0$ if the output of that neuron is negative else returns the output itself. \cite{Gujar11} defines the optimal nonlinear rebate function as the combination of marginal payments. We believe that the rebate functions though nonlinear, should only contain first degree terms for bid values as payments do not have higher order terms, making the function piece-wise linear. Hence, we use ReLU as our activation function.

As required by \Cref{thm:order}, a rebate function for an agent $i$ should depend only on valuations of the remaining agents. Hence, we connect the $i^{th}$ output node to the $i^{th}$ layer of hidden nodes which are connected to all the input nodes except the $i^{th}$ input node as shown in \Cref{fig:Nonlinear}. The redistribution function being anonymous, the weights of the connections entering each hidden layer and each output are the same. The green thick lines in  \Cref{fig:Nonlinear} represent an unique set of weights which are connected to the second agents' output. The same weights are used for calculating the rebate of the other agents as well. The first set of weights ($w$) connect the input nodes to the hidden nodes and bias ($b$) is added to the hidden nodes. The second set of weights ($w'$) connect the hidden nodes to the output nodes and same bias added ($b'$) to each output node.\\ 
\centerline
{$r_i$ = $\displaystyle \sum_{k=1}^{h} relu(\sum_{j=1}^{n-1} v_{i}w_{jk} + b) w'_{k} + b'  $,  $\forall i = 1\mbox{ }to\mbox{ }n,$ } \centerline{$h$: Number of hidden neurons, $relu(x)= max(0,x)$} \\

\tikzset{%
   neuron missing/.style={
    draw=none, 
    scale=4,
    text height=0.333cm,
    execute at begin node=\color{black}$\vdots$
  },
}
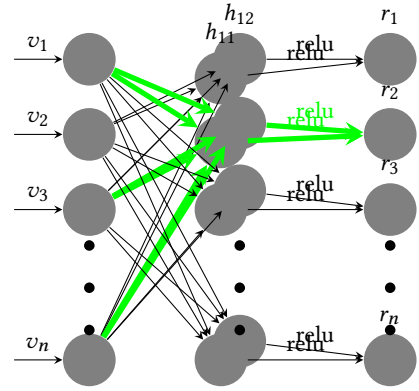
\begin{figure}
\centering
\begin{tikzpicture}[ >=stealth]

\foreach \m/\l [count=\y] in {1,2,3}
{
 \node [circle,fill=black!50,minimum size=0.7cm] (input-\m) at (0,2.5-\y) {};
}
\foreach \m/\l [count=\y] in {4}
{
 \node [circle, fill=black!50,minimum size=0.7cm ] (input-\m) at (0,-2.5) {};
}
 
 \node [neuron missing]  at (0,-1.5) {};

 \foreach \m/\l [count=\y] in {1,2}
{
 \node [circle,fill=black!50,minimum size=0.7cm] (hidden-\m) at (1.5+\y/4,1.0+\y/4) {};
}
 \foreach \m/\l [count=\y] in {3,4}
{
 \node [circle,fill=black!50,minimum size=0.7cm] (hidden-\m) at (1.5+\y/4,0.15+\y/4) {};
}
 \foreach \m/\l [count=\y] in {5,6}
{
 \node [circle,fill=black!50,minimum size=0.7cm] (hidden-\m) at (1.5+\y/4,-0.75+\y/4) {};
}

 \foreach \m/\l [count=\y] in {7,8}
{
 \node [circle,fill=black!50,minimum size=0.7cm] (hidden-\m) at (1.5+\y/4,-2.75+\y/4) {};
}
 
 \node [neuron missing]  at (2,-1.5) {};
 
\foreach \m/\l [count=\y] in {1,2,3}
{
 \node [circle,fill=black!50,minimum size=0.7cm] (output-\m) at (4,2.5-\y) {};
}
\foreach \m/\l [count=\y] in {4}
{
 \node [circle,fill=black!50,minimum size=0.7cm ] (output-\m) at (4,-2.5) {};
}

 \node [neuron missing]  at (4,-1.5) {};
 
\foreach \l [count=\i] in {1,2,3,n}
  \draw [<-] (input-\i) -- ++(-1,0)
    node [above, midway] {$v_{\l}$};
    
\foreach \l [count=\i] in {1,2}
  \node [above] at (hidden-\i.north) {$h_{1\l}$};

\foreach \l [count=\i] in {1,2,3,n}
  \node [above] at (output-\i.north) {$r_{\l}$};

\foreach \i in {1,2}
  \foreach \j in {2,...,4}
    \draw [<-] (hidden-\i) -- (input-\j);

\foreach \i in {3,4}
  \foreach \j in {1,3,4}
    \draw [<-,green,line width=0.70mm] (hidden-\i) -- (input-\j);

\foreach \i in {5,6}
  \foreach \j in {1,2,4}
    \draw [<-] (hidden-\i) -- (input-\j);

\foreach \i in {7,8}
  \foreach \j in {1,2,3}
    \draw [<-] (hidden-\i) -- (input-\j);
    
 \foreach \i in {3,4}
    \draw [<-,green,line width=0.70mm] (output-2) -- (hidden-\i)
    	node[above,midway]{relu};

\foreach \i in {1,2}
    \draw [<-] (output-1) -- (hidden-\i)
    	node[above,midway]{relu};

\foreach \i in {5,6}
    \draw [<-] (output-3) -- (hidden-\i)
    	node[above,midway]{relu};

\foreach \i in {7,8}
    \draw [<-] (output-4) -- (hidden-\i)
    	node[above,midway]{relu};

\end{tikzpicture}
\caption{\label{fig:Nonlinear} Nonlinear Network model}
\end{figure}

The defined network architectures can model different functions depending on the weights. Hence, the training of the network guides the network to learn appropriate weights. Prior to training, we must recall \Cref{thm:order} which necessitates the ordering of the input valuations. Besides the ordering, the constraints defined in \Cref{subsec:bs} require the evaluation of $t$ the VCG Payment.  In the following section, we mention the details about the same.
\subsection{Ordering of Inputs and Payments}
The ordering and calculation of VCG payment in both homogeneous and heterogeneous cases are independent of the neural network. 
\subsubsection{Homogeneous Objects :} \label{sec:homo}
All the given $p$ objects are similar and each agent desires at most one object. The bids submitted are $\theta$ where, $\theta \in \mathbb{R}^{n}$. We order the bids such that $v_{1} \geq v_{2} \geq \ldots \geq v_{n}$. Payment by agent $i$,
\[ t_{i} = \begin{cases} 
      v_{p+1} & i \leq p \\
      0 & i > p 
   \end{cases}
\] Hence, $t = pv_{p+1}$.

\subsubsection{Heterogeneous Objects :} \label{sec:hetero}
All the $p$ objects are different, each agent will submit his valuation for each of the objects. The bids submitted are $\theta$ where $\theta \in \mathbb{R}^{p \times n}$. We define a particular ordering among these vectors based on the overall utility of each agent and the marginal valuations they have for each item. The allocation of the goods is similar to a weighted graph matching problem and is solved using Hungarian Algorithm. Once we get the allocation say, $k^{\ast}$, we proceed to calculate the payments using the VCG payment $t = \sum_{i \in N}t_{i}$, where each $t_{i}$ is given by \Cref{eq:v_pay}. The ordering of bids for the winning $p$ agents is determined based on their utilities. The utility $u_{i}$ of agent $i$ is given by,
$$u_{i} = \displaystyle \sum_{j \in N } v_{j}(k_{*}(\theta),\theta_{j}) - \displaystyle \sum_{j \neq i } v_{j}(k_{-i}^*(\theta_{-i}),\theta_{j}) \mbox{ },\mbox{ } \forall i = 1,\ldots,n. $$ If two agents have same utility, their ordering is determined by their marginal values for the first item, and if it is same, then by second item and so on. Once their ordering is determined, we remove the $p$ agents and then run the VCG mechanism to get the next $p$ winning agents and calculate the ordering using the same procedure as above. If the remaining agents are less than $p$ we can still find the allocation and hence order the remaining agents till none are left or one is left. The time complexity of this ordering is polynomial in $np$. \\

With the given ordering of the inputs and payments, we use the specified network models in both homogeneous and heterogeneous settings. For each setting, we model either OE or OW mechanism. For both OE and OW the network architecture remains same whereas the objective changes as defined in the following section.

\subsection{Objective Function}
During the forward pass the input valuations are multiplied by the network weights and the corresponding rebate for each agent is calculated. The initial weights being random, the rebate calculated will not be optimal. In order to adjust the weights to obtain optimal rebate function, we add an objective at the end of the network which maximizes the rebate in both OW and OE. The loss function essentially is the negative total rebate of all the agents. The objective also takes case of the Feasibility \ref{cond:F} and Individual Rationality condition \ref{cond:IR} 
\subsubsection{Optimal in Expectation (OE)}
\begin{itemize}
\item Given that we need to maximize the total expected rebate, the loss is defined as: $$l(w,b): \frac{1}{T}\displaystyle \sum_{j=1}^{T} \sum_{i=1}^{n} -r_{i}^{j}, $$  $T = $total number of training samples\\
\item Given Inequality constraint for Feasibility \ref{cond:F} we modify it to equality as: 
$$G^{j}(w,b) = max(-g(w,b),0),\mbox{ } \forall j = 1,2...,T$$
\item The overall loss function: \\
\begin{equation}
\label{eq:oe}
L(w,b) = l(w, b) +\frac{\rho}{2}\displaystyle \sum_{j=1}^{T} G^{j}(w,b)^2
\end{equation}
\end{itemize}
\subsubsection{Worst case Optimal (OW)}
\begin{itemize}
\item Given that in OW we are trying to maximize the worst possible redistribution index, as defined in \Cref{def:RI}, the loss is given by: $$l(k): -k $$ 
$$such\mbox{ }that\mbox{ }g_{3}: \displaystyle \sum_{i=1}^n r_{i} - kt \geq 0$$
$T = $ total number of training samples\\
\item  
Given inequality constraint for Feasibility \ref{cond:F} we modify it to equality as:
\[G_{1}^j(w,b) = max(-g(w,b),0)\]
Given inequality constraint for IR \ref{cond:IR} we modify it to equality as: 
\[G_{2}^j(w,b) = max(-g'(w,b),0)\]
The inequality condition for finding the worst case optimal is modified as follows
\[G_{3}^j(w,b) = max(-g_{3}(w,b),0)\] \\
$\forall j = 1,2...,T$
\item The overall loss function for worst case: \\
\begin{equation}
\label{eq:ow}
L(w,b, k) = l(k) +\frac{\rho}{2}\displaystyle \sum_{j=1}^T [G_{1}^j(w,b)^2 + G_{2}^j(w,b)^2 + G_{3}^j(w,b)^2]
\end{equation}
\end{itemize}

The network and objective together can be used to model any mechanism which is either OW or OE.  Taking the cue from \Cref{subsec:opt} we conduct few experiments in order to learn an optimal mechanism which was analytically solved in theory for homogeneous settings. Further, our experiments for heterogeneous settings with objective OW, try to model HETERO \cite{Gujar11} and also OE nonlinear mechanisms for homogeneous setting. 

\section{Implementation Details and Experimental Analysis}
\label{sec:expr}
The proper training of neural networks is very crucial for its convergence. Xavier initialization and Adam optimization guide us in choosing appropriate initialization and optimizer which are crucial for stabilizing the network. Given that we have two different networks and two different objectives, we experimented on various combinations of these to validate the data-driven approach with existing results. In the following subsections, we specify the implementation details.
\subsection{Initialization and Optimizer}
\subsubsection{Xavier initialization:}
\label{sec:xavier}
\cite{xavier10}
The right way of initialization of a neural network is to have weights that are able to produce outputs that follow a similar distribution across all neurons. This will greatly help convergence during training and we will be able to train faster and effectively. Xavier initialization tries to scale the random normal initialized weights with a factor $\alpha$, such that there is unit variance in the output. $\alpha = \frac{1}{\sqrt[]{n}}$, where $n$ is the number of input connection entering in that particular node. $\alpha = \sqrt[]{\frac{2}{n}}$ for ReLU.
\subsubsection{Adam optimizer:}
\label{sec:opt}
\cite{KingmaB14}
Adam is a first order gradient-based optimization of stochastic objective functions. The method computes individual adaptive learning rates for different parameters from estimates of first and second moments of the gradients. The default values provided in the tensorflow library is used for $beta1=0.9,\mbox{ } beta2=0.999\mbox{ }and\mbox{ }epsilon=1e-08.$. The learning rates are different for different cases as mentioned in their respective sections.

\begin{table}
	\caption{$e^{oe}$ for homogeneous and heterogeneous setting.} 
	\label{tab:oe}
  	\begin{tabular}{|l|l|l|l|l|l|}
		\hline
        $n, p$ & \multicolumn{3}{c|}{Homogeneous} & \multicolumn{2}{c|}{Heterogeneous} \\
		\cline{2-6} & {OE-HO} & {OE-HO} &  {OE-HO } & {OE-HE} & {OE-HE}  \\
        & {Theoretical} & {Linear}& {Nonlinear}  & {Linear}& {Nonlinear} \\
        &&NN&NN&NN&NN \\
        \hline
        3,1 & 0.667 & 0.668 & 0.835 & 0.667 & 0.835  \\
        4,1 & 0.833 & 0.836 & 0.916 & 0.834 & 0.920  \\
        5,1 & 0.899 & 0.901 & 0.961 & 0.900 & 0.969  \\
        6,1 & 0.933 & 0.933 & 0.973 & 0.934 & 0.970  \\
        3,2 & 0.667 & 0.665 & 0.839 & 0.458 & 0.774  \\
        4,2 & 0.625 & 0.626 & 0.862 & 0.637 & 0.855  \\
        5,2 & 0.800 & 0.802 & 0.897 & 0.727 & 0.930  \\
        6,2 & 0.875 & 0.875 & 0.935 & 0.756 & 0.954  \\
        10,1 & 0.995 & 0.996 & 0.995 & 0.995 & 0.995 \\
        10,3 & 0.943 & 0.945 & 0.976 & 0.779 & 0.923 \\
        10,5 & 0.880 & 0.880 & 0.947 & 0.791 & 0.897 \\
        10,7 & 0.943 & 0.944 & 0.976 & 0.781 & 0.857 \\
        10,9 & 0.995 & 0.997 & 0.996 & 0.681 & 0.720 \\
 	\hline
	\end{tabular}
\end{table}
\subsection{Different Settings for Training}
\subsubsection{Optimal in Expectation for Homogeneous Objects (OE-HO):} \label{OE_HO}
The inputs form a matrix of $(S \times n)$, where $S$ is the batch size, the values are sampled from a uniform random distribution $~ U[0,1]$. The batch size is set to be as large as possible, for $n<10$, $S = 10000$ and $n=10$, $S = 50000$. After that we apply the ordering and calculate payments for the given input valuations as defined in Section \ref{sec:homo}. Next, we feed it to the linear network model (\Cref{fig:linear}) whose parameters are initialized using Xavier initialization (\Cref{sec:xavier}). The objective function given by \Cref{eq:oe} is applied to the output of the network and parameters are updated using Adam optimizer (\Cref{sec:opt}), learning rate set to $0.0001$. The nonlinear model (\Cref{fig:Nonlinear}) is also trained in the similar manner. We used 1000 nodes in the hidden layer and the network was trained with a learning rate of $10e-4$.   
\subsubsection{Optimal in Worst-case for Homogeneous Objects (OW-HO):} \label{OW_HO}
As in the OE case, linear network model is used and similar procedure is followed. The input is sorted as given in \Cref{sec:homo} and along with the calculated payments is fed to the linear network. The objective given in \Cref{eq:ow} is optimized with the learning rate set to $0.0001$ and the training is carried till the loss decreases and saturates which happens when the redistribution index is optimal. As discussed in \Cref{subsec:wco} for homogeneous setting linear rebate functions are optimal among all possible deterministic functions which are DSIC and AE, hence we did not use nonlinear model for this case.
\subsubsection{Optimal in Expectation for Heterogeneous Objects (OE-HE):} \label{OE_HE}
The inputs are again randomly sampled from a uniform distribution $~U[0,1]$, the input matrix is of the form, $(S \times n \times p)$. Then the inputs are ordered as defined in \Cref{sec:hetero}. Both the networks \Cref{fig:linear}, \Cref{fig:Nonlinear} are used for finding the optimal in expectation mechanism. Just like in the homogeneous case, network parameters are initialized using Xavier initialization. For the payment calculation, we use the scipy library for linear sum assignment. This library assigns objects such that the cost is minimized, whereas we want the valuation to be maximized as per AE, hence we negate the bids before passing it to the function. Besides, the Hungarian algorithm for assignment works only when the number of objects to be assigned is same as the agents, hence we introduce dummy agents or dummy objects with zero valuation so that the input matrix is a square matrix.  The objective function \ref{eq:oe} is optimized using Adam optimizer (described in \Cref{sec:opt}) with learning rate $10e-4$ for both the linear and nonlinear models. In the nonlinear network, the number of nodes in the hidden layer was set to $1000$.  

\subsubsection{Optimal in Worst-case for Heterogeneous Objects (OW-HE):} \label{OW_HE}
For designing this particular mechanism, we use the same inputs that we used in the OE setting for heterogeneous items. The networks used and their initialization is also same. The only difference is the objective function which is given by \Cref{eq:ow} and the optimizer used is Adam. For the linear network learning rate is $10e-4$. In the nonlinear network, the number of hidden nodes used were $100$ and learning rate of $10e-4$ for all values of $n$, with $p=1$. When the values of $p >1$, the number of hidden nodes was increased to 1000 and a learning rate of $10e-5$ was used.
\noindent

\begin{table}
\caption{$e^{ow}$ for Homogeneous and Heterogeneous setting.} 
\label{tab:ow}
\begin{tabular}{|l|l|l|l|l|}
\hline
$n,p$ & \multicolumn{2}{c|}{Homogeneous }& \multicolumn{2}{c|}{Heterogeneous} \\
\cline{2-5} & {OW-HO} & {OW-HO} & {OW-HE} \footnotemark & {OW-HE } \\
&{ theoretical} & {Linear NN} & {Linear NN} & {Nonlinear NN} \\
\hline
3,1 & 0.333 & 0.336 & 0.332 & 0.333  \\
4,1 & 0.571 & 0.575 & 0.571 & 0.571 \\
4,2 & 0.250 & 0.250 & 0.0 & 0.250 \\
5,1 & 0.733 & 0.739 & 0.733 & 0.732 \\
5,2 & 0.454 & 0.460 & 0.0 & 0.454 \\
5,3 & 0.200 & 0.200 & 0.0 & 0.199 \\
6,1 & 0.839 & 0.847 & 0.839& 0.838  \\
6,2 & 0.615 & 0.620 & 0.0 & 0.614  \\
6,3 & 0.375 & 0.378 & 0.0 & 0.375 \\ 
7,1 & 0.905 & 0.910 & 0.905 & 0.904 \\
7,2 & 0.737 & 0.746 & 0.0 & 0.736  \\
7,3 & 0.524 & 0.538 & 0.0  & 0.523  \\
8,1 & 0.945 & 0.949 & 0.945  & 0.943  \\
8,2 & 0.825 & 0.834 & 0.0 & 0.825 \\
9,1 & 0.969 & 0.972 & 0.968 & 0.968 \\
9,2 & 0.887 & 0.894 & 0.0 & 0.886  \\
10,1 & 0.982 & 0.985 & 0.982 & 0.982  \\
10,2 & 0.928 & 0.936 & 0.0 & 0.927  \\
\hline
\end{tabular}
\end{table}
\footnotetext{All values below 10e-3 are considered to be 0.0}

\begin{figure*}
\captionsetup{font=scriptsize}
\label{graphs}
\centering
\begin{minipage}{.33\linewidth}
  \includegraphics[width=1.0\columnwidth]{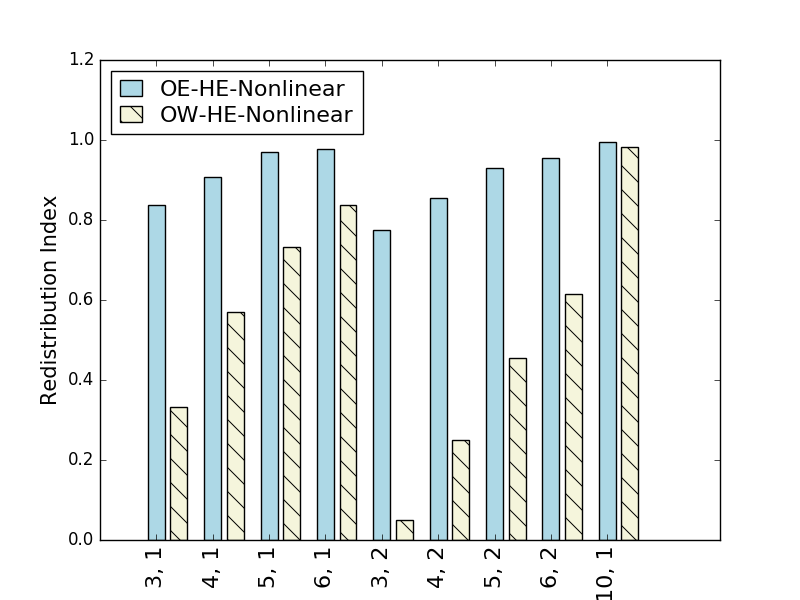}
  \captionof{figure}{OE-HE-Nonlinear Vs OW-HE-Nonlinear}
  \label{img1}
\end{minipage}
\begin{minipage}{0.33\linewidth}
  \includegraphics[width=\linewidth]{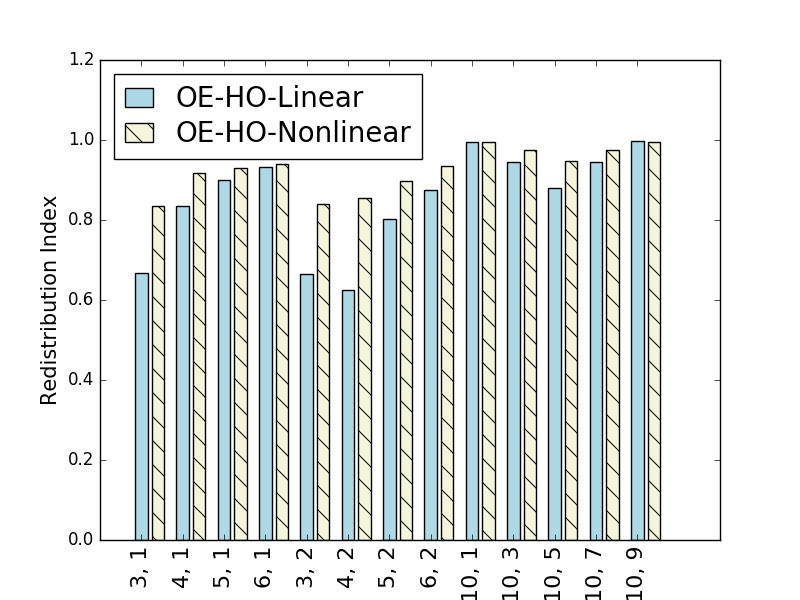}
  \captionof{figure}{OE-HO-Linear vs OE-HO-Nonlinear}
  \label{img2}
\end{minipage}
\begin{minipage}{0.33\linewidth}
\includegraphics[width=1.0\columnwidth]{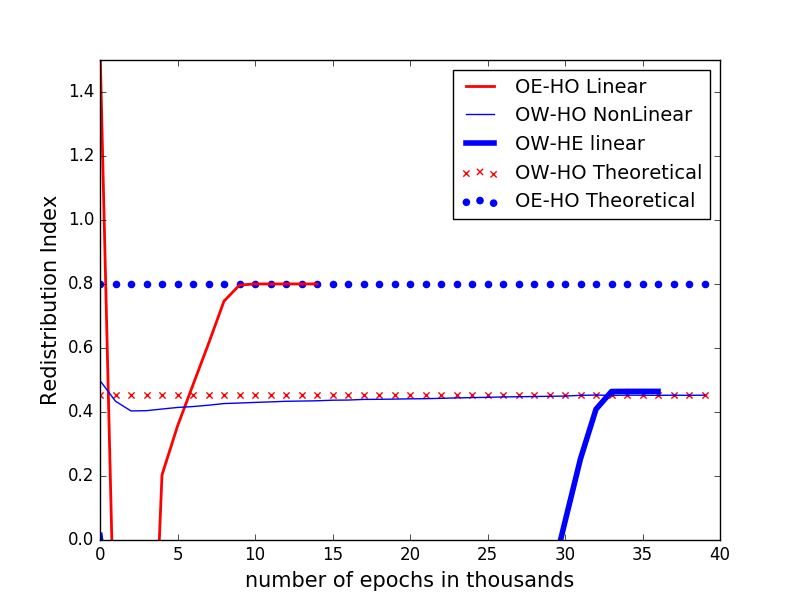} 
\caption{RI values with change \\ in epoch for $n = 5$, $p = 2$}
\label{img3}
\end{minipage}
\end{figure*}
\subsection{Specific parameters used for training}
\begin{itemize}
\item We used tensorflow library for all the implementations. We used GPUs: Tesla K40c and GeForce GTX Titan X. The network training time  varies from few minutes to a whole day, for the different $n,\mbox{ }p$ values. 
\item The number of input samples, $T$ for binary settings ideally should be $2^{np}$ and way more than this for the real value cases. So we use 10000 samples where $np \leq 13 $, 70000 for $np \leq 16$ and $100000$ for the rest of the cases.
\item All the experiments are run for a maximum of 400000 epochs and the constant $\rho$ as given in the overall loss functions (\ref{eq:oe}, \ref{eq:ow}) is set to 1000.
\item The number of nodes in the hidden layer for the network given in \Cref{fig:Nonlinear} is 1000 ideally. Even having 100 is sufficient for the cases where the product of n and p is less than 10.
\end{itemize}

\subsection{Results and Discussion}
We now describe the results obtained from the networks trained as discussed above. \Cref{tab:oe} compares the values of redistribution index ($e^{oe}$) for all the experiments with OE objective for different ($n$,$p$) values under homogeneous and heterogeneous settings. The first column indicates theoretical bounds on OE RMs with linear rebate functions. In the column  heterogeneous, we compare the redistribution indexes obtained by our OE-HE networks with linear and nonlinear networks. Similarly, \Cref{tab:ow} compares the OW redistribution index ($e^{ow}$) for all the networks under consideration and theoretical values for different  ($n$,$p$) values. With these tables, the following are our observations:

\emph{Achieving the analytically solved bounds:} 
For the OE objective in homogeneous setting, our network OE-HO-Linear achieves the theoretical values of  $e^{oe}$ proposed in \cite{GuoOEL08}. Similarly, the theoretical $e^{ow}$ values are achieved by the network OW-HO Linear.

\emph{OW nonlinear rebate function for heterogeneous setting :}
The values from OE-HE-Linear NN illustrate the impossibility theorem stated in \cite{Gujar11} that there cannot be a linear rebate function with non-zero redistribution index for heterogeneous settings. For $p=1$, there being no difference in homogeneous or heterogeneous, the values remain the same, but are zero for $p>1$. The network OW-HE-Nonlinear achieves the theoretical values given in OW-HO theoretical \Cref{tab:ow}. 

\emph{OE nonlinear rebate function for homogeneous setting :} \cite{GuoOEL08} have only tried to find the OE linear rebate function. OE-HO Nonlinear NN outperforms the linear counterpart. The graph in \Cref{img1} illustrates the comparison. This indicates the existence of a nonlinear rebate function which guarantees higher $e^{oe}$ than the linear rebate function for homogeneous setting. 

\emph{OE nonlinear rebate function for heterogeneous setting :}
 The values from OE-HE-Linear NN is shows same results as OE-HO Linear for $p=1$ and different otherwise. OE-HE-Nonlinear NN outperforms the linear network The graph in \Cref{img2} compares between the OW and OE performance of the nonlinear network for heterogeneous settings.
 
\Cref{img1} illustrates the significantly better performance of Optimal in Expectation RM wrt Optimal in Worst Case for heterogeneous settings for different $(n,p)$ values. For reference, it also has OE performance for homogeneous settings. We also observe that performance of OE mechanisms with nonlinear rebates is significantly better as compared to OE mechanism with linear rebates (\Cref{img2}). 
The convergence of neural networks for $n=5,p=2$ with number of epochs while training can be found in \Cref{img3}. Typically, most of the networks studied here converge in less than 80000 epochs for $n\leq 10$. (Whenever the objective value of a neural network drops below zero, we skip them to plot).

\section{Conclusions}
\label{sec:con}
In this paper, we considered a problem of designing an optimal redistribution mechanism. We proposed a novel approach to this problem: train a neural network with randomly generated valuation profiles for a desired goal. Our neural network design takes care of problem specific constraints. We showed that, it can learn optimal redistribution mechanisms with  proper initialization and a suitably defined ordering over valuation profiles.
Our analysis shows that one can design  nonlinear rebate functions for homogeneous settings that perform better than optimal in expectation linear rebate functions.  We could design optimal in expectation rebate functions for heterogeneous objects which is not solved analytically. 

There are many challenges to be handled here. For example, can we design a vanilla  neural network that can learn linear, nonlinear rebate functions without explicitly designing such architectures, based on the problem specific needs? Can we come up with training strategies independent of $(n,p)$?

\pagebreak
\bibliographystyle{ACM-Reference-Format}  
\bibliography{references}  


\begin{thebibliography}{00}


\ifx \showCODEN    \undefined \def \showCODEN     #1{\unskip}     \fi
\ifx \showDOI      \undefined \def \showDOI       #1{{\tt DOI:}\penalty0{#1}\ }
  \fi
\ifx \showISBNx    \undefined \def \showISBNx     #1{\unskip}     \fi
\ifx \showISBNxiii \undefined \def \showISBNxiii  #1{\unskip}     \fi
\ifx \showISSN     \undefined \def \showISSN      #1{\unskip}     \fi
\ifx \showLCCN     \undefined \def \showLCCN      #1{\unskip}     \fi
\ifx \shownote     \undefined \def \shownote      #1{#1}          \fi
\ifx \showarticletitle \undefined \def \showarticletitle #1{#1}   \fi
\ifx \showURL      \undefined \def \showURL       #1{#1}          \fi
\providecommand\bibfield[2]{#2}
\providecommand\bibinfo[2]{#2}
\providecommand\natexlab[1]{#1}

\bibitem[\protect\citeauthoryear{Bailey}{Bailey}{1997}]%
        {Bailey97}
\bibfield{author}{\bibinfo{person}{Martin~J Bailey}.}
  \bibinfo{year}{1997}\natexlab{}.
\newblock \showarticletitle{The Demand Revealing Process: To Distribute the
  Surplus}.
\newblock \bibinfo{journal}{{\em Public Choice\/}} \bibinfo{volume}{{91}, 2}
  (\bibinfo{year}{1997}), \bibinfo{pages}{107--26}.
\newblock
\showURL{%
\url{https://EconPapers.repec.org/RePEc:kap:pubcho:v:91:y:1997:i:2:p:107-26}}


\bibitem[\protect\citeauthoryear{Cavallo}{Cavallo}{2006}]%
        {cavalloRedis06}
\bibfield{author}{\bibinfo{person}{Ruggiero Cavallo}.}
  \bibinfo{year}{2006}\natexlab{}.
\newblock \showarticletitle{Optimal Decision-Making With Minimal Waste:
  Strategyproof Redistribution of VCG Payments}. In \bibinfo{booktitle}{{\em
  Proc. of the 5th Int. Joint Conf. on Autonomous Agents and Multi Agent
  Systems (AAMAS{\textquoteright}06)}}. Hakodate, Japan,
  \bibinfo{pages}{882--889}.
\newblock
\showURL{%
\url{http://econcs.seas.harvard.edu/files/econcs/files/cavallo-redis.pdf}}


\bibitem[\protect\citeauthoryear{Clarke}{Clarke}{1971}]%
        {Clarke71}
\bibfield{author}{\bibinfo{person}{Edward Clarke}.}
  \bibinfo{year}{1971}\natexlab{}.
\newblock \showarticletitle{Multipart pricing of public goods}.
\newblock \bibinfo{journal}{{\em Public Choice\/}} \bibinfo{volume}{{11}, 1}
  (\bibinfo{year}{1971}), \bibinfo{pages}{17--33}.
\newblock
\showURL{%
\url{https://EconPapers.repec.org/RePEc:kap:pubcho:v:11:y:1971:i:1:p:17-33}}


\bibitem[\protect\citeauthoryear{de~Clippel, Naroditskiy, Polukarov, Greenwald,
  and Jennings}{de~Clippel et~al\mbox{.}}{2014}]%
        {Clippel14}
\bibfield{author}{\bibinfo{person}{Geoffroy de Clippel},
  \bibinfo{person}{Victor Naroditskiy}, \bibinfo{person}{Maria Polukarov},
  \bibinfo{person}{Amy Greenwald}, {and} \bibinfo{person}{Nicholas~R.
  Jennings}.} \bibinfo{year}{2014}\natexlab{}.
\newblock \showarticletitle{Destroy to save}.
\newblock \bibinfo{journal}{{\em Games and Economic Behavior\/}}
  \bibinfo{volume}{{86}, C} (\bibinfo{year}{2014}), \bibinfo{pages}{392--404}.
\newblock


\bibitem[\protect\citeauthoryear{D{\"{u}}tting, Feng, Narasimhan, and
  Parkes}{D{\"{u}}tting et~al\mbox{.}}{2017}]%
        {optAuctions}
\bibfield{author}{\bibinfo{person}{Paul D{\"{u}}tting}, \bibinfo{person}{Zhe
  Feng}, \bibinfo{person}{Harikrishna Narasimhan}, {and}
  \bibinfo{person}{David~C. Parkes}.} \bibinfo{year}{2017}\natexlab{}.
\newblock \showarticletitle{Optimal Auctions through Deep Learning}.
\newblock \bibinfo{journal}{{\em CoRR\/}}  \bibinfo{volume}{abs/1706.03459}
  (\bibinfo{year}{2017}).
\newblock
\showURL{%
\url{http://arxiv.org/abs/1706.03459}}


\bibitem[\protect\citeauthoryear{Faltings}{Faltings}{2005}]%
        {Faltings05}
\bibfield{author}{\bibinfo{person}{Boi Faltings}.}
  \bibinfo{year}{2005}\natexlab{}.
\newblock \showarticletitle{A Budget-balanced, Incentive-compatible Scheme for
  Social Choice}. In \bibinfo{booktitle}{{\em Proceedings of the 6th AAMAS
  International Conference on Agent-Mediated Electronic Commerce: Theories for
  and Engineering of Distributed Mechanisms and Systems}} \bibinfo{series}{{\em
  (AAMAS'04)}}. Springer-Verlag, Berlin, Heidelberg, \bibinfo{pages}{30--43}.
\newblock
\showISBNx{3-540-29737-5, 978-3-540-29737-6}
\showDOI{%
\url{http://dx.doi.org/10.1007/11575726_3}}


\bibitem[\protect\citeauthoryear{Garg, Narahari, and Gujar}{Garg
  et~al\mbox{.}}{2008}]%
        {Garg08a}
\bibfield{author}{\bibinfo{person}{Dinesh Garg}, \bibinfo{person}{Y Narahari},
  {and} \bibinfo{person}{Sujit Gujar}.} \bibinfo{year}{2008}\natexlab{}.
\newblock \showarticletitle{Foundations of Mechanism Design: A Tutorial - Part
  1: Key Concepts and Classical Results}.
\newblock \bibinfo{journal}{{\em Sadhana - Indian Academy Proceedings in
  Engineering Sciences\/}} \bibinfo{volume}{{33}, Part 2}
  (\bibinfo{date}{April} \bibinfo{year}{2008}), \bibinfo{pages}{83--130}.
\newblock


\bibitem[\protect\citeauthoryear{Glorot and Bengio}{Glorot and Bengio}{2010}]%
        {xavier10}
\bibfield{author}{\bibinfo{person}{Xavier Glorot} {and} \bibinfo{person}{Yoshua
  Bengio}.} \bibinfo{year}{2010}\natexlab{}.
\newblock \showarticletitle{Understanding the difficulty of training deep
  feedforward neural networks}. In \bibinfo{booktitle}{{\em Proceedings of the
  Thirteenth International Conference on Artificial Intelligence and
  Statistics}} \bibinfo{series}{{\em (Proceedings of Machine Learning
  Research)}}, \bibfield{editor}{\bibinfo{person}{Yee~Whye Teh} {and}
  \bibinfo{person}{Mike Titterington}} (Eds.), \bibinfo{volume}{Vol.~9}. PMLR,
  Chia Laguna Resort, Sardinia, Italy, \bibinfo{pages}{249--256}.
\newblock
\showURL{%
\url{http://proceedings.mlr.press/v9/glorot10a.html}}


\bibitem[\protect\citeauthoryear{Graves}{Graves}{2013}]%
        {Graves13}
\bibfield{author}{\bibinfo{person}{Alex Graves}.}
  \bibinfo{year}{2013}\natexlab{}.
\newblock \showarticletitle{Generating Sequences With Recurrent Neural
  Networks}.
\newblock \bibinfo{journal}{{\em CoRR\/}}  \bibinfo{volume}{abs/1308.0850}
  (\bibinfo{year}{2013}).
\newblock


\bibitem[\protect\citeauthoryear{Green and Laffont}{Green and Laffont}{1979}]%
        {Green81}
\bibfield{author}{\bibinfo{person}{Jerry~R. Green} {and}
  \bibinfo{person}{Jean-Jacques Laffont}.} \bibinfo{year}{1979}\natexlab{}.
\newblock \bibinfo{booktitle}{{\em Incentives in Public Decision Making}}.
\newblock North-Holland, Amsterdam.
\newblock


\bibitem[\protect\citeauthoryear{Groves}{Groves}{1973}]%
        {Groves73}
\bibfield{author}{\bibinfo{person}{Theodore Groves}.}
  \bibinfo{year}{1973}\natexlab{}.
\newblock \showarticletitle{Incentives in Teams}.
\newblock \bibinfo{journal}{{\em Econometrica\/}} \bibinfo{volume}{{41}, 4}
  (\bibinfo{year}{1973}), \bibinfo{pages}{617--31}.
\newblock
\showURL{%
\url{https://EconPapers.repec.org/RePEc:ecm:emetrp:v:41:y:1973:i:4:p:617-31}}


\bibitem[\protect\citeauthoryear{Gujar and Narahari}{Gujar and
  Narahari}{2011}]%
        {Gujar11}
\bibfield{author}{\bibinfo{person}{Sujit Gujar} {and} \bibinfo{person}{Y.
  Narahari}.} \bibinfo{year}{2011}\natexlab{}.
\newblock \showarticletitle{Redistribution Mechanisms for Assignment of
  Heterogeneous Objects}.
\newblock \bibinfo{journal}{{\em J. Artif. Int. Res.\/}} \bibinfo{volume}{{41},
  2} (\bibinfo{date}{May} \bibinfo{year}{2011}), \bibinfo{pages}{131--154}.
\newblock
\showISSN{1076-9757}
\showURL{%
\url{http://dl.acm.org/citation.cfm?id=2051237.2051242}}


\bibitem[\protect\citeauthoryear{Gul and Stacchetti}{Gul and
  Stacchetti}{1999}]%
        {Gul99}
\bibfield{author}{\bibinfo{person}{Faruk Gul} {and} \bibinfo{person}{Ennio
  Stacchetti}.} \bibinfo{year}{1999}\natexlab{}.
\newblock \showarticletitle{Walrasian Equilibrium with Gross Substitutes}.
\newblock \bibinfo{journal}{{\em Journal of Economic Theory\/}}
  \bibinfo{volume}{{87}, 1} (\bibinfo{year}{1999}), \bibinfo{pages}{95--124}.
\newblock
\showURL{%
\url{https://EconPapers.repec.org/RePEc:eee:jetheo:v:87:y:1999:i:1:p:95-124}}


\bibitem[\protect\citeauthoryear{Guo}{Guo}{2011}]%
        {Guo11}
\bibfield{author}{\bibinfo{person}{Mingyu Guo}.}
  \bibinfo{year}{2011}\natexlab{}.
\newblock \bibinfo{title}{VCG Redistribution with Gross Substitutes}.
\newblock   (\bibinfo{year}{2011}).
\newblock


\bibitem[\protect\citeauthoryear{Guo}{Guo}{2012}]%
        {Guo12}
\bibfield{author}{\bibinfo{person}{Mingyu Guo}.}
  \bibinfo{year}{2012}\natexlab{}.
\newblock \showarticletitle{Worst-case Optimal Redistribution of VCG Payments
  in Heterogeneous-item Auctions with Unit Demand}. In \bibinfo{booktitle}{{\em
  Proceedings of the 11th International Conference on Autonomous Agents and
  Multiagent Systems - Volume 2}} \bibinfo{series}{{\em (AAMAS '12)}}.
  International Foundation for Autonomous Agents and Multiagent Systems,
  Richland, SC, \bibinfo{pages}{745--752}.
\newblock
\showISBNx{0-9817381-2-5, 978-0-9817381-2-3}
\showURL{%
\url{http://dl.acm.org/citation.cfm?id=2343776.2343803}}


\bibitem[\protect\citeauthoryear{Guo and Conitzer}{Guo and Conitzer}{2007}]%
        {Guo07}
\bibfield{author}{\bibinfo{person}{Mingyu Guo} {and} \bibinfo{person}{Vincent
  Conitzer}.} \bibinfo{year}{2007}\natexlab{}.
\newblock \showarticletitle{Worst-case Optimal Redistribution of VCG Payments}.
  In \bibinfo{booktitle}{{\em Proceedings of the 8th ACM Conference on
  Electronic Commerce}} \bibinfo{series}{{\em (EC '07)}}. ACM, New York, NY,
  USA, \bibinfo{pages}{30--39}.
\newblock
\showISBNx{978-1-59593-653-0}
\showDOI{%
\url{http://dx.doi.org/10.1145/1250910.1250915}}


\bibitem[\protect\citeauthoryear{Guo and Conitzer}{Guo and Conitzer}{2008a}]%
        {Guo08}
\bibfield{author}{\bibinfo{person}{Mingyu Guo} {and} \bibinfo{person}{Vincent
  Conitzer}.} \bibinfo{year}{2008}\natexlab{a}.
\newblock \showarticletitle{Better Redistribution with Inefficient Allocation
  in Multi-unit Auctions with Unit Demand}. In \bibinfo{booktitle}{{\em
  Proceedings of the 9th ACM Conference on Electronic Commerce}}
  \bibinfo{series}{{\em (EC '08)}}. ACM, New York, NY, USA,
  \bibinfo{pages}{210--219}.
\newblock
\showISBNx{978-1-60558-169-9}
\showDOI{%
\url{http://dx.doi.org/10.1145/1386790.1386825}}


\bibitem[\protect\citeauthoryear{Guo and Conitzer}{Guo and Conitzer}{2008b}]%
        {GuoOEL08}
\bibfield{author}{\bibinfo{person}{Mingyu Guo} {and} \bibinfo{person}{Vincent
  Conitzer}.} \bibinfo{year}{2008}\natexlab{b}.
\newblock \showarticletitle{Optimal-in-expectation Redistribution Mechanisms}.
  In \bibinfo{booktitle}{{\em Proceedings of the 7th International Joint
  Conference on Autonomous Agents and Multiagent Systems - Volume 2}}
  \bibinfo{series}{{\em (AAMAS '08)}}. International Foundation for Autonomous
  Agents and Multiagent Systems, Richland, SC, \bibinfo{pages}{1047--1054}.
\newblock
\showISBNx{978-0-9817381-1-6}
\showURL{%
\url{http://dl.acm.org/citation.cfm?id=1402298.1402367}}


\bibitem[\protect\citeauthoryear{Guo and Conitzer}{Guo and Conitzer}{2009}]%
        {Guo09}
\bibfield{author}{\bibinfo{person}{Mingyu Guo} {and} \bibinfo{person}{Vincent
  Conitzer}.} \bibinfo{year}{2009}\natexlab{}.
\newblock \showarticletitle{Worst-case optimal redistribution of VCG payments
  in multi-unit auctions}.
\newblock \bibinfo{journal}{{\em Games and Economic Behavior\/}}
  \bibinfo{volume}{{67}, 1} (\bibinfo{year}{2009}), \bibinfo{pages}{69--98}.
\newblock


\bibitem[\protect\citeauthoryear{Hartline and Roughgarden}{Hartline and
  Roughgarden}{2008}]%
        {Hartline08}
\bibfield{author}{\bibinfo{person}{Jason~D. Hartline} {and}
  \bibinfo{person}{Tim Roughgarden}.} \bibinfo{year}{2008}\natexlab{}.
\newblock \showarticletitle{Optimal Mechanism Design and Money Burning}. In
  \bibinfo{booktitle}{{\em Proceedings of the Fortieth Annual ACM Symposium on
  Theory of Computing}} \bibinfo{series}{{\em (STOC '08)}}. ACM, New York, NY,
  USA, \bibinfo{pages}{75--84}.
\newblock
\showISBNx{978-1-60558-047-0}
\showDOI{%
\url{http://dx.doi.org/10.1145/1374376.1374390}}


\bibitem[\protect\citeauthoryear{He, Zhang, Ren, and Sun}{He
  et~al\mbox{.}}{2016}]%
        {kaiming16}
\bibfield{author}{\bibinfo{person}{Kaiming He}, \bibinfo{person}{Xiangyu
  Zhang}, \bibinfo{person}{Shaoqing Ren}, {and} \bibinfo{person}{Jian Sun}.}
  \bibinfo{year}{2016}\natexlab{}.
\newblock \showarticletitle{Identity Mappings in Deep Residual Networks}.
\newblock \bibinfo{journal}{{\em CoRR\/}}  \bibinfo{volume}{abs/1603.05027}
  (\bibinfo{year}{2016}).
\newblock
\showURL{%
\url{http://arxiv.org/abs/1603.05027}}


\bibitem[\protect\citeauthoryear{Hornik}{Hornik}{1991}]%
        {hornik91}
\bibfield{author}{\bibinfo{person}{Kurt Hornik}.}
  \bibinfo{year}{1991}\natexlab{}.
\newblock \showarticletitle{Approximation Capabilities of Multilayer
  Feedforward Networks}.
\newblock \bibinfo{journal}{{\em Neural Networks\/}}  \bibinfo{volume}{4}
  (\bibinfo{year}{1991}), \bibinfo{pages}{251--257}.
\newblock
\showDOI{%
\url{http://dx.doi.org/10.1016/0893-6080(91)90009-T}}


\bibitem[\protect\citeauthoryear{Hornik, Stinchcombe, and White}{Hornik
  et~al\mbox{.}}{1989}]%
        {hornik89}
\bibfield{author}{\bibinfo{person}{Kurt Hornik}, \bibinfo{person}{Maxwell
  Stinchcombe}, {and} \bibinfo{person}{Halbert White}.}
  \bibinfo{year}{1989}\natexlab{}.
\newblock \showarticletitle{Multilayer feedforward networks are universal
  approximators}.
\newblock \bibinfo{journal}{{\em Neural networks\/}} \bibinfo{volume}{{2}, 5}
  (\bibinfo{year}{1989}), \bibinfo{pages}{359--366}.
\newblock


\bibitem[\protect\citeauthoryear{Kawaguchi}{Kawaguchi}{2016}]%
        {Kawaguchi16}
\bibfield{author}{\bibinfo{person}{Kenji Kawaguchi}.}
  \bibinfo{year}{2016}\natexlab{}.
\newblock \showarticletitle{Deep Learning without Poor Local Minima}.
\newblock In \bibinfo{booktitle}{{\em Advances in Neural Information Processing
  Systems 29}}, \bibfield{editor}{\bibinfo{person}{D.~D. Lee},
  \bibinfo{person}{M.~Sugiyama}, \bibinfo{person}{U.~V. Luxburg},
  \bibinfo{person}{I.~Guyon}, {and} \bibinfo{person}{R.~Garnett}} (Eds.).
  Curran Associates, Inc., \bibinfo{pages}{586--594}.
\newblock
\showURL{%
\url{http://papers.nips.cc/paper/6112-deep-learning-without-poor-local-minima.pdf}}


\bibitem[\protect\citeauthoryear{Kingma and Ba}{Kingma and Ba}{2014}]%
        {KingmaB14}
\bibfield{author}{\bibinfo{person}{Diederik~P. Kingma} {and}
  \bibinfo{person}{Jimmy Ba}.} \bibinfo{year}{2014}\natexlab{}.
\newblock \showarticletitle{Adam: {A} Method for Stochastic Optimization}.
\newblock \bibinfo{journal}{{\em CoRR\/}}  \bibinfo{volume}{abs/1412.6980}
  (\bibinfo{year}{2014}).
\newblock
\showURL{%
\url{http://arxiv.org/abs/1412.6980}}


\bibitem[\protect\citeauthoryear{Maskin, Laffont, and Laffont}{Maskin
  et~al\mbox{.}}{1979}]%
        {LaffontM79}
\bibfield{author}{\bibinfo{person}{Eric Maskin}, \bibinfo{person}{J.~J.
  Laffont}, {and} \bibinfo{person}{J.~J. Laffont}.}
  \bibinfo{year}{1979}\natexlab{}.
\newblock \bibinfo{booktitle}{{\em A Differential Approach to Expected Utility
  Maximizing Mechanisms}}.
\newblock North Holland, 289-308.
\newblock


\bibitem[\protect\citeauthoryear{Moulin}{Moulin}{2009}]%
        {Moulin09}
\bibfield{author}{\bibinfo{person}{Herve Moulin}.}
  \bibinfo{year}{2009}\natexlab{}.
\newblock \showarticletitle{Almost budget-balanced VCG mechanisms to assign
  multiple objects}.
\newblock \bibinfo{journal}{{\em Journal of Economic Theory\/}}
  \bibinfo{volume}{{144}, 1} (\bibinfo{year}{2009}), \bibinfo{pages}{96--119}.
\newblock


\bibitem[\protect\citeauthoryear{Nisan}{Nisan}{2007}]%
        {nisan07}
\bibfield{author}{\bibinfo{person}{Noam Nisan}.}
  \bibinfo{year}{2007}\natexlab{}.
\newblock \showarticletitle{Introduction to Mechanism Design (for Computer
  Scientist}.
\newblock In \bibinfo{booktitle}{{\em Algorithmic game theory}},
  \bibfield{editor}{\bibinfo{person}{Noam Nisan}, \bibinfo{person}{Tim
  Roughgarden}, \bibinfo{person}{Eva Tardos}, {and} \bibinfo{person}{Vijay
  Vazirani}} (Eds.). Cambridge University Press, \bibinfo{pages}{209--242}.
\newblock


\bibitem[\protect\citeauthoryear{Parkes, Kalagnanam, and Eso}{Parkes
  et~al\mbox{.}}{2001}]%
        {parkes01}
\bibfield{author}{\bibinfo{person}{David~C. Parkes}, \bibinfo{person}{Jayant~R.
  Kalagnanam}, {and} \bibinfo{person}{Marta Eso}.}
  \bibinfo{year}{2001}\natexlab{}.
\newblock \showarticletitle{Achieving Budget-Balance with Vickrey-Based Payment
  Schemes in Exchanges}. In \bibinfo{booktitle}{{\em Proc. 17th International
  Joint Conference on Artificial Intelligence (IJCAI{\textquoteright}01)}}.
  \bibinfo{pages}{1161{\textendash}1168}.
\newblock
\showURL{%
\url{http://econcs.seas.harvard.edu/files/econcs/files/combexch01.pdf}}


\bibitem[\protect\citeauthoryear{Soudry and Carmon}{Soudry and Carmon}{2016}]%
        {daniel16}
\bibfield{author}{\bibinfo{person}{Daniel Soudry} {and} \bibinfo{person}{Yair
  Carmon}.} \bibinfo{year}{2016}\natexlab{}.
\newblock \showarticletitle{No bad local minima: Data independent training
  error guarantees for multilayer neural networks}.
\newblock \bibinfo{journal}{{\em CoRR\/}}  \bibinfo{volume}{abs/1605.08361}
  (\bibinfo{year}{2016}).
\newblock
\showURL{%
\url{http://arxiv.org/abs/1605.08361}}


\bibitem[\protect\citeauthoryear{Tang}{Tang}{2017}]%
        {tang17}
\bibfield{author}{\bibinfo{person}{Pingzhong Tang}.}
  \bibinfo{year}{2017}\natexlab{}.
\newblock \showarticletitle{Reinforcement mechanism design}. In
  \bibinfo{booktitle}{{\em Proceedings of the Twenty-Sixth International Joint
  Conference on Artificial Intelligence, {IJCAI-17}}}.
  \bibinfo{pages}{5146--5150}.
\newblock
\showDOI{%
\url{http://dx.doi.org/10.24963/ijcai.2017/739}}


\bibitem[\protect\citeauthoryear{Vickrey}{Vickrey}{1961}]%
        {Vickrey61}
\bibfield{author}{\bibinfo{person}{W. Vickrey}.}
  \bibinfo{year}{1961}\natexlab{}.
\newblock \showarticletitle{Counterspeculation, Auctions, and Competitive
  Sealed Tenders}.
\newblock \bibinfo{journal}{{\em Journal of Finance\/}} \bibinfo{volume}{{16},
  1} (\bibinfo{date}{March} \bibinfo{year}{1961}), \bibinfo{pages}{8--37}.
\newblock


\end{thebibliography}

\end{document}